# Stationary Algorithmic Balancing
# For Dynamic Email Re-Ranking Problem

Jiayi Liu
liu2861@purdue.edu
Department of Computer Science
Purdue University
West Lafayette, Indiana, United States

Jennifer Neville
neville@purdue.edu
Department of Computer Science
Purdue University / Microsoft Research
West Lafayette, Indiana, United States

## ABSTRACT

Email platforms need to generate personalized rankings of emails that satisfy user preferences, which may vary over time. We approach this as a recommendation problem based on three criteria: closeness (how relevant the sender and topic are to the user), timeliness (how recent the email is), and conciseness (how brief the email is). We propose MOSR (Multi-Objective Stationary Recommender), a novel online algorithm that uses an adaptive control model to dynamically balance these criteria and adapt to preference changes. We evaluate MOSR on the Enron Email Dataset, a large collection of real emails, and compare it with other baselines. The results show that MOSR achieves better performance, especially under non-stationary preferences, where users value different criteria more or less over time. We also test MOSR's robustness on a smaller down-sampled dataset that exhibits high variance in email characteristics, and show that it maintains stable rankings across different samples. Our work offers novel insights into how to design email re-ranking systems that account for multiple objectives impacting user satisfaction.

## CCS CONCEPTS

• **Information systems** → **Data streaming**; **Data stream mining**; **Email**; **Rank aggregation**; **Learning to rank**; **Recommender systems**; **Social networks**.

## KEYWORDS

objective balancing, online recommendation system

## 1 INTRODUCTION

Email is one of the most popular online activities, with millions of users exchanging messages every day. However, managing a large and diverse email inbox can be overwhelming and frustrating for users, reducing their satisfaction and productivity [6, 26]. Therefore, designing email platforms that can help users cope with email overload and find the most important messages to send or reply is a key challenge. Email recommender systems aim to provide personalized suggestions for ranking emails based on users' preferences [14]. For example, Google's 'Priority Inbox' feature ranks emails according to their inferred priority for reading based on users' past behavior [1].

However, user preferences are not static; they may change over time depending on various factors such as context or mood. To account for this dynamic nature of preferences, email recommender systems need to learn from feedback and update their ranking strategies accordingly. Offline methods that assume fixed or stable preferences may fail to capture the evolving interests of users over time [15, 20]. Thus, an online algorithm that can adapt to preference changes in real time is crucial.

Moreover, email recommendation is not a single-objective problem; it involves multiple criteria that affect user satisfaction with different aspects of emails. In this paper we focus on three criteria: closeness (how relevant the sender and topic are to the user), timeliness (how urgent the email is), conciseness (how brief the email is). These criteria reflect different dimensions of importance that users may value differently at different times. For instance, a user may prefer timely but concise emails during busy workdays but close but lengthy ones during leisure time. Hence, email recommender systems need to balance these multiple objectives while generating personalized rankings.

Existing approaches for email re-ranking or recommendation have mostly focused on maximizing relevance or priority based on certain features. For example, some methods use sender-receiver relationship features [2, 7], others use topic models [1, 4, 13, 30], while others combine text similarity with temporal features [8]. However, these methods have some limitations in terms of accuracy and adaptability. They neglect other factors besides relevance such as novelty or diversity which may also influence user satisfaction [20] . Most importantly they do not explicitly account for preference changes over time. Recent research has started considering "beyond relevance" objectives in recommendation systems, such as exploration vs exploitation, serendipity vs familiarity etc., which optimize factors affecting user engagement rather than just item relevance [3, 5, 20]. We argue that similar objectives apply to email ranking settings, where users may value different aspects of emails more or less at different times depending on their context.

In this paper, we address this problem as a multi-objective online recommendation task based on three criteria: closeness, timeliness,

and conciseness. Closeness refers to the estimation of the relationship between the sender and receiver, timeliness refers to the urgency of a reply, and conciseness refers to the usage of words in the email. We argue that these aspects reflect different dimensions of user satisfaction with respect to emails, and they may vary across different users and over time.

We propose MOSR (Multi-Objective Stationary Recommender), a novel online algorithm that uses an adaptive control model to balance these criteria and adapt to preference changes. Our algorithm learns each criterion's weight from historical data and updates it using gradient descent based on observed feedback signals. It then combines these weights into a single score for each email using a linear aggregation function. By doing so, our algorithm can adjust its ranking strategy according to changing preferences without requiring prior knowledge or explicit input from users.

The main contributions of our work are as follows:

(1) We formulate the email re-ranking as a multi-objective online recommendation problem that aims to optimize three criteria: closeness, timeliness, and conciseness. These are key factors that influence user actions in email. We show how preferences w.r.t these criteria vary across users and over time.
(2) We propose MOSR , an adaptive control model that learns a reference vector from historical data and adjusts it based on online feedback. The reference vector represents the relative importance of each criterion for each user at each moment. MOSR adapts the reference vector dynamically by using reinforcement learning techniques without requiring re-training or compromising privacy.
(3) We evaluate MOSR on the Enron Email Dataset[16]. We show that MOSR outperforms several baselines in terms of ranking quality measured by NDCG. We also demonstrate that MOSR handles non-stationary preferences well by providing consistent recommendations even when users change their values for different criteria over time. Furthermore, we test MOSR 's robustness on a smaller dataset sampled randomly at different time intervals and show that MOSR still performs better than other methods under high variance conditions.

## 2 MOSR FRAMEWORK

Our goal is to design a recommendation system that helps users choose when and how to send emails based on their preferences w.r.t. relationships, urgency, and brevity. We model this as a dynamic problem that involves multiple objectives that may conflict or change over time. Our algorithm uses the email stream flow as input and tries to find the optimal trade-offs among these objectives for each email ranking decision.

### 2.1 Problem Definition

Our re-ranking problem is a type of recommendation problem that consists of two stages: candidate generation and ranking. However, it differs from the typical recommendation problem in two ways:

- First, we need to balance multiple and sometimes conflicting criteria to achieve the highest level of satisfaction among them.

- Second, users' email ranking preferences are not fixed but may change depending on external factors. Figure 2 illustrates some scenarios where users' preferences vary or remain constant due to different influences.

*Definition 2.1 (Email object).* We consider an email object consists to be represented as $G = \{s, u, c, t\}$, where $s \in E$ is the email address of the sender, $u \in E$ is the email address of the receivers, $c$ is the content of the email, $t$ is the timestamp when $G$ is sent. **G** is the set of all the email objects.

We want to rank the emails of a specific email address $e_i$ according to the user's preferences, which may change over time. The ranking candidates are the emails that have been received or sent by $e_i$.

*Definition 2.2 (Candidate Set).* Here, we define candidates set $Q$ with $Q = \{q_1, q_2, ..q_n\}$. There are two types of candidates in $Q$: unanswered emails in the inbox or follow-up emails after no response. Hence, the candidates set $Q$ includes the people $e_i$ sent to/received from. As we defined before, candidates set $Q = \{e_1, e_2, ..e_j, ...\}, e_j \in E$. At different timestamps $t_i$, the candidate set would also be updated with time window $t_w$.

Note the candidate set $Q$ is not fixed, since new emails may arrive or be sent at any time. To rank the candidates, we assign each email a score based on multiple criteria $\Phi, \Xi, \Upsilon$ for the current timestamp $t_i$. These criteria reflect how relevant, timely, important, or interesting an email is for the user. We also use a feedback-based aggregation function that can adjust the scores online as we learn from different users' choices. Then we sort the emails by their scores to get a personalized ranking for each user at any time.

*Definition 2.3 (Loss function).* We define our candidates set $Q$ as a set of emails, $Q = \{e_1, e_2, ...\}$, and our prediction $y$ as the ranking of our candidates. Then, for a given algorithm $\Omega$, the predicted ranking **y** could be defined as $\mathbf{y} = [p^\Omega(e_1), p^\Omega(e_2), ...]^T$, in which $p^\Omega(e_k)$ represents the predicted ranking of $e_k$. Suppose the predicted score for a candidate $e$ is $\Omega(e)$, then $p^\Omega(e) = k | \Omega(e) = \Omega(\mathbf{e})_{D(k)}$. Here, $\Omega(\mathbf{e})$ is the vector of predicted scores for **e** under algorithm $\Omega$, and $\Omega(\mathbf{e})_{D(k)}$ follows Definition 3.1.

### 2.2 Proposed Approach

The overall architecture of our algorithm is depicted in Figure 1b. In this section, we will introduce the details of the MOSR algorithm.

(1) Step 1: Weighting preferences (RIM+OWA, see Sec 3.3)
(2) Step 2: Identify candidates set $Q$.
(3) Step 3: Rank the candidates $Q$, with weighted function.
(4) Step 4: Compute loss, update the weights with MRAC and repeat.

We use several ordered weight averaging (OWA) aggregators to combine the criteria of closeness, timeliness, and conciseness and obtain the predicted scores of candidates Q. Then, we apply a weighted sum aggregator to re-rank the scores from OWA. To adapt to users' choices, we adjust the weights of different scores by adaptive control over the multi-score aggregation. For each email address $e_i$, we update its sending preference online according to the loss between true ranking and predicted ranking. When $e_i$ sends an email to a candidate $Q_j$, it raises the priority of $Q_j$ and the MRAC (Model Reference Adaptive Control—defined below) modifies the weights of relevant scores.

We formulate our problem as a dynamic multiple objective optimization problem to achieve algorithmic balance over closeness, timeliness, and conciseness. The conventional multiple objective optimization problem aims to optimize the weight of different objectives under constrained or conflicting situations [17]. However, this is not suitable for our case because email history changes over time. Therefore, we propose a dynamic version that involves multi-stage ranking setups and time windows.

Most existing recommendation systems use a two-stage mechanism: they first extract potential candidates and model their features to get one score per candidate [18, 19, 30]. However, this is inefficient for data streams because learning over large candidate sets becomes impractical. Unlike previous systems, we use multiple scores based on different user habits instead of one general static score. We also use time windows to enable fast switching among different scores as email history evolves.

We propose a MRAC (Model Reference Adaptive Control) model to create an online mechanism for the multi-objective optimization problem. In this model, we use different rankers to order the solutions according to various criteria, with the aim of discovering the personalized preferences of each user over these rankers. We assume that there is a true preference ranking that reflects the user's ideal ordering of solutions, and our goal is to estimate and update the user's preference over different rankers as they interact with them. To do this, we treat each ranker as a fixed model, and we measure the distance between the true ranking and the predicted ranking by each ranker as a controller.

## 3 BACKGROUND

### 3.1 Email overloading problem

Many users face the problem of email overload, where their inboxes are filled with too many emails and they struggle to identify or respond to the important ones [6, 26]. One possible solution is to re-rank incoming emails and create a priority inbox based on various factors [1]. Previous studies have explored different aspects of this problem, such as how people decide whether to reply or not, depending on interpersonal differences, email content, attachments, and other features [4, 11, 12, 30]. They also proposed methods to predict the priority of emails in the inbox using content-based features [10, 13, 24].

However, most of these methods rely on analyzing the content of emails, which may raise privacy concerns. Aberdeen et al.[1] used a linear logistic regression model with multiple content-based features for real-time online ranking. Yang et al.[30] included attachments as an additional feature for analysis. Feng et al.[13] developed a doc2vec based generative model to rank inbox emails. Bedekar et al.[4] re-ranked emails according to their topic analysis.

In this work, we examine how different criteria affect email ranking jointly.

### 3.2 Model Reference Adaptive Control

MRAC is a control method that uses a reference system (model) as a target for the process being controlled. The reference system has a model with state, input and output variables. The controller

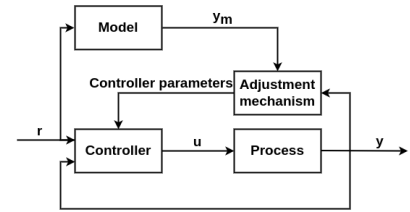

**(a) Model Reference Adaptive Control (MRAC)**

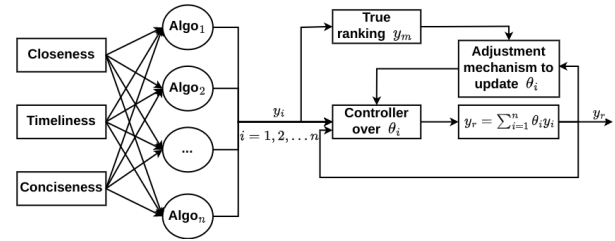

**(b) Architecture of MOSR**

**Figure 1:** Figure (a) shows the flow chart of MRAC, consisting of Reference model, Process model, Controller and Adaption algorithm. Figure (b) shows the architecture of MOSR. The detailed training process is in 5.3.

parameters change in real-time using an adaptive optimization algorithm. Fig 1a shows the main parts of the MRAC: Reference model, Process model, Controller and Adaption algorithm.

*3.2.1 Elements in MRAC.*

- Reference Model
  The reference model defines the desired behavior of a process and is usually expressed in a parametric form (e.g., transfer function/state-space models) that can be implemented in the control computer. To achieve an exact match between the reference model and the actual process, the reference model must have some properties: it must be stable and minimum phase (meaning that its poles/zeros are in the left-half plane), and it must represent the process well.
- Controller
  An MRAC system requires a controller that meets some criteria. First, it must ensure "perfect model matching", which means that there must exist control parameters that make the closed-loop response identical to that of the reference model. Second, it must use direct adaptation, which means that the control parameters depend on a linear function of the error signal. In our model, we use OWA-related algorithms to estimate these control parameters based on minimizing an objective function.

*3.2.2 Adaptive control with multiple fixed models.* MRAC aims to optimize the controller parameters for the entire system. However, some controllers may rely on multiple models in the system [21, 22]. How to switch and tune between models is a common topic. The models can be either fixed or adaptive. A fixed model has constant controller parameters, while an adaptive model requires parameter adjustment. An MRAC algorithm with multiple models should specify how to select the appropriate controller for different environments.

## 3.3 Multi-Objective Optimization

One way to combine multiple criteria into a single decision function is by using ordered weight averaging functions (OWA) [28]. These functions aggregate the scores that measure how well different criteria are satisfied [9]. However, unlike weighted sum functions that assign fixed weights to each criterion, OWA functions assign weights based on the magnitude of the scores. This means that higher scores indicate more important criteria. OWA functions are often used in recommendations that involve several satisfaction criteria, such as music recommendations and COVID-19 policy [20, 23].

*Definition 3.1.* For any vector $\mathbf{x}$, we denote $\mathbf{x} \searrow$ as the vector obtained from $\mathbf{x}$ with a non-increasing order. For simplicity, we name $\mathbf{x} \searrow = \mathbf{x}_D$. Then we have $\mathbf{x}_{D(0)} \geq \mathbf{x}_{D(1)} \geq ... \geq \mathbf{x}_{D(n)}$ [20].

As a symmetric aggregation function, OWA assigns weights according to the values of attributes. Thus, each weight is not associated with a particular attribute. Giving an input $\mathbf{x}$ and a weighting vector $\mathbf{w}$, the OWA function will be

$$OWA_w(\mathbf{w}, \mathbf{x}) = \sum_{i=1}^{n} w_i x'_i \quad (1)$$

where $x'_i$ is $i$-th largest element in $\mathbf{x}$, or to say $\mathbf{x}_{D(i)}$. There are many methods to obtain the weighting vector $\mathbf{w}$. One typical method is the use of Regular Increasing Monotone (RIM) quantifiers [29]. RIM quantifiers generate the weights by

$$w_i = R(\frac{i}{n}) - R(\frac{i-1}{n}) \quad (2)$$

in which $i$ is the $i$-th largest value, $n$ is the number of criteria in OWA, and $R$ is the RIM quantifier. Furthermore, RIM restricts that $\sum_i w_i = 1$. Hence, a typical quantifier is [29]

$$R(x) = x^\alpha \quad \alpha \geq 0 \quad (3)$$

Since $x \in \{\frac{0}{n}, \frac{1}{n}, ... \frac{n}{n}\}$. The changes over parameters $\alpha$ bring the RIM quantifier to different cases. When $\alpha \to 0$, the OWA becomes the MAX operator, when $\alpha \to 0$, the OWA operator becomes the arithmetic mean, and when $\alpha \to \infty$, the OWA operator becomes the MIN operator.

## 4 USER PREFERENCES

Our goal is to understand what makes an email more or less important to users based on closeness, timeliness, and conciseness. Here, we define the criteria we use to assess these factors and quantify them in our recommendation.

*Definition 4.1 (Insider space).* Suppose $S$ is a subset of $E$, indicating the insider email addresses. Here, we define insider email addresses as Enron email addresses, and outsider email as non-Enron email addresses. We define the insider space $\mathbf{I} = S \times \mathbf{G}$ and a function $f$ over $\mathbf{I}$.

*Definition 4.2 (Flow list).* Denote the flow list of each email address $e_i$ in set $S$ as $U_i = [u_1, u_2, \ldots, u_j, \ldots]$. In this representation, each element $u_j$ in the flow list $U_i$ corresponds to an email in the set $\mathbf{G}$, where $u_j$ is either sent or received by the email address $e_i$. We will predict priority for each new item in the flow list to re-rank the emails.

Table 1: Notations used in this paper

| Symbol | Meaning |
| --- | --- |
| $e$ | email address |
| $G$ | email object |
| $U$ | flow list of an email address |
| $l$ | job level for the owner of an email address |
| $t$ | time stamp |
| $t_w$ | time window |
| $c$ | the content of email |
| $Q$ | candidate set |

*Definition 4.3 (Job level).* We introduce a surjective function $\pi$ to map $e_i \in S$ to $L$, in which, $L$ is the set of job levels, where $L = [k]$. For each $e_i \in S$, we have $l_i = \pi(e_i)$. We denote $\pi$ acting on $S$ as $\pi(S) = \{\pi(e_i) | e_i \in S\}$. Note that $\pi(S) = L$.

A flow list is a key concept for online email re-ranking problems because it enables online training updates. Rather than re-train the model with a new large-scale dataset, we can add the new data to the flow list and train them together.

We treat the email re-ranking problem as a recommendation problem that has two stages: candidate generation and ranking. However, our re-ranking problem differs from typical recommendation problems in three main ways:

- The set of candidate emails changes dynamically as new emails arrive or old ones are replied to.
- We need to balance different and sometimes conflicting criteria to achieve optimal satisfaction for the user.
- There is no universal formula to capture all users' email ranking preferences because they may vary depending on the context and mood of the user. Figure 2 illustrates this.

For a given email address $e_i$, we will compute a set of candidates $Q = \{q_1, q_2, ..q_n\}$. Note that $Q$ will change with time flows. Our goal is to score the candidates above with an aggregation function and update the function online with different users' choices. By sorting the scores we get, we will obtain the ranking of candidates.

### 4.1 Key Concepts

In this paper, we use three key concepts: closeness, timeliness, and conciseness. We will define and measure them in the next section.

*4.1.1 Closeness.* Closeness represents the relationship between users. People may prefer to reply to those who are closer to them when they prioritize their emails. We distinguish between insider closeness and outsider closeness. Insider closeness captures the relationship between Enron employees, based on their relative job level. Outsider closeness reflects the relationship between Enron employees and people from other organizations.

**Quantifying closeness:** We adopt two criteria to quantify closeness between two email addresses $e_i$ and $e_j$: (1) Their previous email history frequency, and (2) Their business relationship. To capture dynamics in previous email frequency, we adopt a sliding time window $w_\Phi$. We first define *frequency* $f_{(t_i,t_j)}(e_i, e_j)$ as the number of emails from $e_i$ to $e_j$ between $t_i$ to $t_j$. Suppose $e_i$ is the sender and $e_j$ is the recipient, then at timestamp $t_i$, the to-frequency $f_t$ is $f_{(t_i-w,t_i)}(e_i,e_j)$ and the from-frequency $f_f$ is $f_{(t_i-w,t_i)}(e_j,e_i)$. To quantify their previous email history frequency $\gamma$ at $t_i$, we apply a weighting for the to-frequency and the from-frequency, thus $\gamma_{t_i}(e_i,e_j) = e^{w_1 f_t + w_2 f_f}$.

- Insider closeness. Insider closeness $\Phi_1$ is defined when the sender and receiver are in the same company. In this case, we include a job level ratio in the measure. Let the job levels of sender $e_i$ and recipient $e_j$ be $l_i$ and $l_j$, respectively. Then the insider closeness between them is:

$$\Phi_1(e_i, e_j) = \mathbb{1}[e_i \in E]\mathbb{1}[e_j \in E]\gamma_{t_i}(e_i, e_j) \cdot e^{\frac{l_j}{2*l_i}}$$

- Outsider closeness. Outsider closeness $\Phi_2$ is defined when the receiver is from a different company than the sender:

$$\Phi_2(e_i, e_j) = \mathbb{1}[e_i \in E]\mathbb{1}[e_j \notin E]\gamma_{t_i}(e_i, e_j)d(e_i, e_j)$$

Here, $d(e_i, e_j)$ indicates the social distance between $e_i$ and $e_j$. We will further describe this in the section below.

*4.1.2 Timeliness.* Timeliness indicates how long it has been since the last email. Timeliness affects the urgency of a reply. Usually, people tend to reply sooner to emails they receive earlier[30]. We also account for cases where people send follow-up emails.

**Quantifying timeliness:** Timeliness helps quantify the urgency to reply to an email and, in turn, composes the preference over email replying priority. Suppose for sender $e_i$ and recipient $e_j$, we have the subset of chatting history $h_i$ with only $e_j$; refer to this as $\partial_{e=e_j}(h_i)$, in which $\partial$ is a filter. Then the time stamps for $e_i$ sending emails to $e_j$ will be $t_s = \partial_{e=e_j,d=1}(h_i)$, and the time stamps for $e_i$ receiving emails from $e_j$ will be $t_r = \partial_{e=e_j,d=-1}(h_i)$. Timeliness consists of two aspects: reply timeliness and follow-up timeliness. To account for those two aspects, we estimate a score for each aspect. Suppose the current timestamp is $t_i$, for reply timeliness, the score is $\Xi_r = t_i - max(t_r)$. For follow-up timeliness, the score will be $\Xi_f = t_i - min(w_\Xi, max(t_s))$, in which $w_\Xi$ is the time window for follow-up timeliness. We apply weights to $\Xi_r, \Xi_f$ to form timeliness $\Xi = \alpha_1 \Xi_r + \alpha_2 \Xi_f$

*4.1.3 Conciseness.* Conciseness measures the ratio of useful information in an email.

**Quantifying conciseness:** The ratio of stop-words helps us approximate how much useful information is in an email. Thus we quantify conciseness as the ratio of non-stop-words. Suppose the content is $c_i$ at time $t_i$ with $s_i$ stop-words, then the conciseness is $1 - \frac{len(s_i)}{len(c_i)}$.

## 4.2 Relationships Between Multiple Objectives

The heatmaps in Figure 2 illustrate how different criterion relate to each other for different users. To construct these heatmaps, we consider seven distinct scores, namely insider score, outsider score, length of email content, effective length ratio of email content, receiving time, number of 2-paths, and their corresponding replying priority ranking. These scores are calculated for each email within the flow list denoted as $U_i$. Subsequently, we compute the correlation coefficient between the aforementioned objectives for each user denoted as $e_i$.

The graphs show how user behaviors changed before and after the Enron Scandal. We analyzed how different types of users prioritized replying emails of different types over time. For example, Kenneth Lay, the CEO of Enron, replied more quickly to outsider emails after the scandal broke out. On the other hand, Kim Ward, a Trader Manager, delayed responding to outsider emails. Marie Heard, an Enron lawyer, maintained a similar pattern of replies before and after the scandal. However, her boss, Stephanie Panus, responded less promptly to insider emails.

Our analysis reveals two noteworthy patterns: First, the attributes we use vary in how much they reveal and how well they match different users' needs, so there is no one-size-fits-all model for user behavior. Second, the relationships among attributes can shift over time depending on external events. For example, after the Enron Scandal, some users kept their email habits unchanged, while others changed their preferences drastically.

These observations motivate our approach. First, since the attributes represent different aspects of satisfaction that may conflict with each other, simply adding them up with weights will not work well. Therefore, we need a flexible model that can adapt to each user's situation. Second, we assume that unpredictable factors can affect users' preferences significantly over time. Hence, an online algorithm that can adjust to changing conditions is essential for re-ranking emails effectively.

Based on these observations, we develop an MRAC-based symmetric aggregation method to address these challenges.

## 5 MOSR DETAILS

### 5.1 Candidate Set Construction

*5.1.1 Graph construction and social distance calculation.* In outsider closeness, we introduce distance $d(e_i, e_j)$, which is the social distance between $e_i$ and $e_j$. To calculate social distance, we need to first establish a social network graph between email addresses.

*Definition 5.1.* For two email address $e_i$ and $e_j$, suppose the number of emails $e_i$ sent to $e_j$ is $n_{e_i \to e_j}$, while the number of emails $e_j$ sent to $e_i$ is $n_{e_j \to e_i}$. If $n_{e_i \to e_j}, n_{e_j \to e_i}$ satisfy the pre-defined restriction, we could establish an edge between $e_i$ and $e_j$. Note that, there would be more edges over time.

$$\begin{cases} n_{e_i \to e_j} \geq k_1, \\ n_{e_j \to e_i} \geq k_2, \\ n_{e_i \to e_j} + n_{e_j \to e_i} \geq k_3, \end{cases} \quad (4)$$

Here, $k_1, k_2, k_3$ are the parameters we will use in experiments.

After computing the graph between email addresses, we will measure the social distance $d(e_i)$. There are two options for the distance measurements for . We adopt two options to calculate the social $d(e_i, e_j)$. The first one is the shortest distance between $e_i$ and $e_j$, the second one is the number of 2-paths between $e_i$ and $e_j$[27]. We will further discuss these two options in the experiment part. In our experiment, the parameters we adopt are $[k_1, k_2, k_3] = [0, 0, 2]$.

Suppose there is an email that is sent by $s_j$ to $u_j$ at time $t_j$, then we will have:

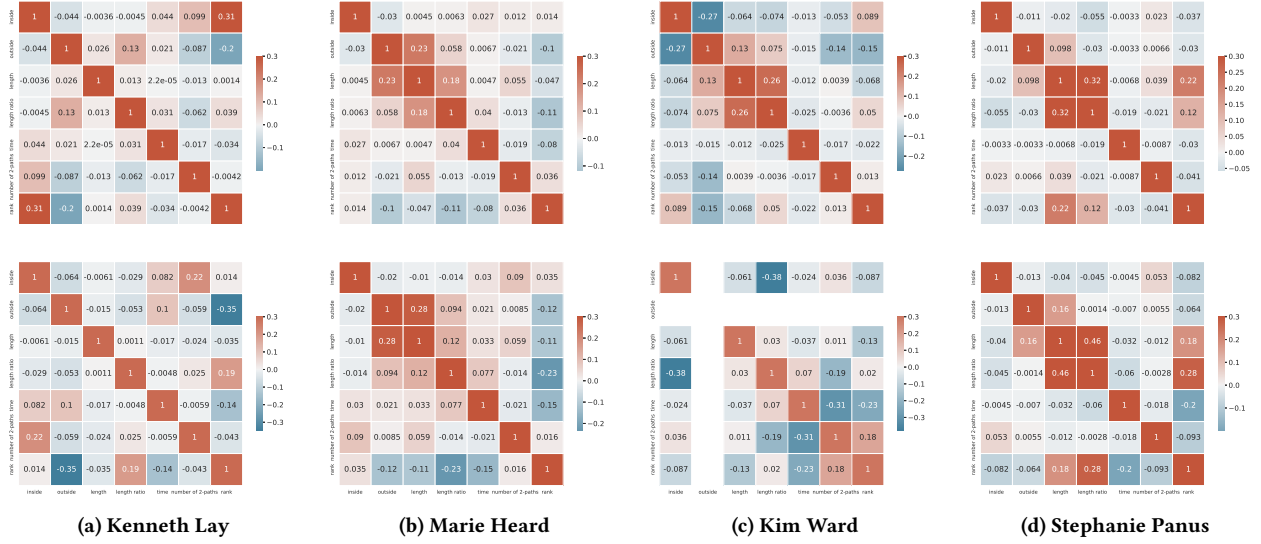

Figure 2: The upper graphs are the user behaviors before Enron Scandal happened, while the below graphs are after Enron Scandal happened.

- Unanswered emails in the inbox. At timestamp $t_j$, $q_j \in E$ will be added to the candidates set of $u_j$, name it $Q_{u_j}$. If $u_j$ sends $t_j$ and email in $[t_j, t_j + t_w]$ or the current time stamp reaches $t_j + t_w$, it will be removed from $Q_{u_j}$.
- Follow-up emails after no response. Name the candidates set of $s_j$ as $Q_{s_j}$. **On the second day** after $t_j$, $u_j$ will be added to $Q_{s_j}$. If $u_j$ sends $s_j$ email in $[t_j, t_j + t_w]$ or the current time stamp reaches $t_j + t_w$, it will be removed from $Q_{s_j}$.

We compare the predicted candidate set $Q$ to the true candidate set $\hat{Q}$ of emails sent by the users. Since the candidate set may not cover the whole true set, we further analyze the composition of the undiscovered candidates in $\hat{Q} - Q$. We found that around 35.5% of these undiscovered candidates have mutual connections with the sender. However, as the graph grows larger, adding more mutual connections as candidates will lower the precision. We also explore the impact of adding carbon copy recipients to the candidate set but found that it did not have any significant effect. The results of analysis over mutual connections are included in the Appendix for further exploration.

### 5.2 Loss

Our prediction process consists of two steps: generating a candidate set and re-ranking it. Given that our predicted candidate set may not align precisely with the scope of the ground truth, we calculate how much our prediction $y$ deviates from the true value $y_m$ by applying methods below:

*Definition 5.2.* We define our candidates set $Q$ as a set of emails, $Q = \{e_1, e_2, ...\}$, and our prediction $y$ as the ranking of our candidates. Then, for a given algorithm $\Omega$, the predicted ranking $\mathbf{y}$ could be defined as $\mathbf{y} = [p^\Omega(e_1), p^\Omega(e_2), ...]^T$, in which, $p^\Omega(e_k)$ represents the predicted ranking of $e_k$. Suppose the predicted score for a candidate $e$ is $\Omega(e)$, then $p^\Omega(e) = k | \Omega(e) = \Omega(\mathbf{e})_{D(k)}$. Here,

$\Omega(\mathbf{e})$ is the vector of predicted scores for $\mathbf{e}$ under algorithm $\Omega$, and $\Omega(\mathbf{e})_{D(k)}$ follows Definition 3.1.

*Definition 5.3.* Suppose our predicted candidates set and results are $Q$ and $y$, the ground truth candidates set and results are $\hat{Q}$ and $y_m$. Then for algorithm $\Omega$, the loss between $\mathbf{y}$ and $\mathbf{y_m}$ comes from two parts: the difference between ranking of discovered candidates $Q \cap \hat{Q}$, and undiscovered candidates $\hat{Q} - Q$. Then we define loss function as

$$\epsilon_r = ||\mathbf{y_r} - \mathbf{y_m}||_2 \\ = \big(\sum_{e \in Q \cap \hat{Q},} (p_r^\Omega(e) - p_m^\Omega(e))^2 + \sum_{e \in \hat{Q} - Q} \delta_d^2\big)^{\frac{1}{2}} \quad (5)$$

Here $\delta_d$ is the parameter we use to measure the cover rate of our predicted candidates set. We define $\mathbf{y_r}$ as the ranking results of MOSR and $\mathbf{y_1}, \mathbf{y_2}, ... \mathbf{y_n}$ as the ranking results of algorithms $\Omega_1, \Omega_2, ... \Omega_n$.

When $\delta_d = 0$, then the loss between $y_r$ and $y_m$ will only consider their ranking. When $\sigma$ grows larger, the importance of cover rate on candidates will be larger.

### 5.3 Training Process

To generate a ranking with email flows, we use the following steps:
(1) Step 1: Use RIM quantifier $R$ to generate OWA weights and construct multiple OWA operators $\Omega_i$.
(2) Step 2: Construct a weighted sum aggregator for the generated OWA operators. The weight for $\Omega_i$ is $\theta_i$, $\sum_{i=1}^n \theta_i = 1$.
(3) Step 3: Construct graph and update candidates set $Q$.
(4) Step 4: Calculate the score and rank the candidates, with score $\mathbf{y_r} = \sum_{i=1}^n \theta_i \mathbf{y_i}$.
(5) Step 5: Compute the loss of results in step 4, update the weights with MRAC and go back to step 3.

To reduce the loss with MRAC, we adjust the weights for each OWA operator. Unlike traditional machine learning methods, MRAC aims to match the current system and minimize the gap between predicted output and actual output. We think that email reply preferences may vary and are not fixed. Therefore, we need an algorithm that can track all the user's dynamics, so that it can continuously optimize the parameters to adapt to uncertainty.

Suppose in $j$-iteration, the loss of $y_i$ is $\epsilon_i$, with $\mathcal{E} = [\epsilon_1, \epsilon_2, ...\epsilon_n]$. Then we will update $\theta_i$ by

$$\theta_i^{(j+1)} = \begin{cases} \frac{\lambda \theta_i^{(j)} + 1}{\lambda + 1} & \text{if } \epsilon_i = \min(\mathcal{E}) \\ \frac{\lambda \theta_i^{(j)}}{\lambda + 1} & \text{otherwise} \end{cases} \quad (6)$$

Here $\lambda$ is the learning rate.

Theorem 5.4. *Equation 5.3 satisfies Lyapunov-stable.*

The proof of Theorem 5.4 is in Appendix.

## 6 EXPERIMENTS

### 6.1 Dataset

In this paper, we use the Enron email dataset, which consists of 500K emails sent by 1K employees of the Enron Corporation. To better understand their priority preferences, we extract the job titles of 200 key employees from the web. We filter the dataset to include only emails from 1999 to 2002 and show its statistics in Figure3. We organize the dataset as a time flow to mimic a real-world email recommendation system. Since the Enron scandal broke out in October 2001, we divide the dataset into two training sets: EnronA and EnronB. EnronA covers January 1999 to March 2001 for training and April 2001 to December 2001 for testing. EnronB covers January 2001 to July 2001 for training and August.

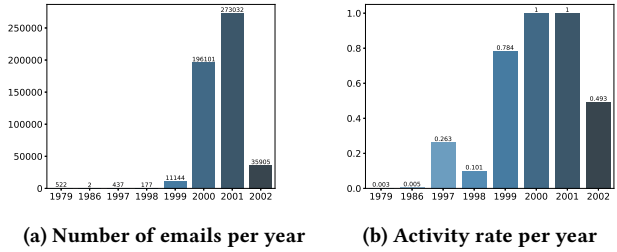

(a) Number of emails per year  (b) Activity rate per year

Figure 3: Email activity in the Enron Dataset by year. In (a), we see how many emails were sent and received each year. The data for 1999 is incomplete, so we exclude it from our analysis. In (b), we see the continuity of email activity each year. This is the percentage of days in a year when at least one email was sent or received. The activity rate was 100% in 2000 and 2001, indicating daily email communication.

### 6.2 Methods Compared

We evaluate different baselines for ranking email messages: Logistic Regression (MS-LR), AdaBoost (MS-ADA)[30], four rankers based on ordered weighted averaging (OWA), a simple time-based ranker, and our proposed method MOSR. Table 2 shows the average loss of each method, which measures how well they match user preferences. The OWA-based rankers have lower losses than MS-LR and MS-ADA, confirming that symmetric aggregation is better than asymmetric weighting. Furthermore, MOSR achieves the lowest loss among all methods and adapts to changing user preferences during re-ranking.

Figure 4-5 compare the daily losses of our method and the baselines. The figures show a noticeable increase in loss for the baselines that use MSFT features in late 2001, when the Enron scandal broke out. This matches the observed shift in sentiment towards Enron at that time, as reported by [25]. Figure 4-5 also reveal the variability of user preferences over time. In later sections, we will show how our algorithm MOSR can cope with such variations and maintain a stable performance regardless of external factors affecting user preferences. We provide full results in Appendix due to space limitations.

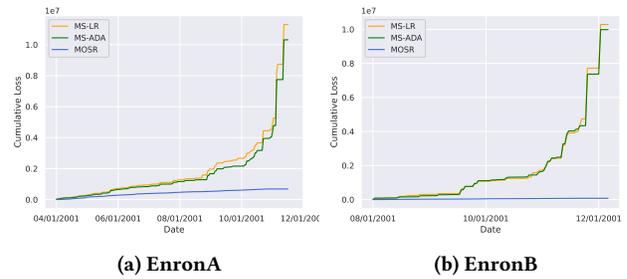

(a) EnronA  (b) EnronB

Figure 4: Cumulative loss curve by date on EnronA and EnronB. This curve represents the sum of losses that occurred from the start of the period until a given date. Both datasets show steeper slopes in their curves after October 2001, which means that losses grew faster in the subsequent months. This coincides with the time of the Enron Scandal.

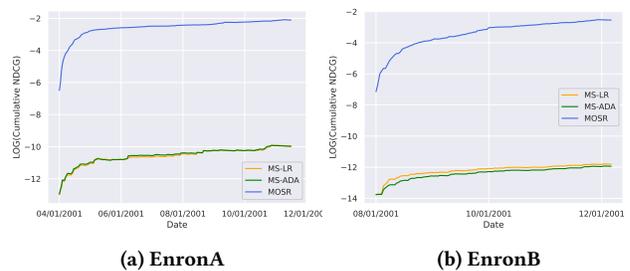

(a) EnronA  (b) EnronB

Figure 5: Logarithm of the sum of the NDCG scores for each day for EnronA and EnronB. The curve flattens out after October 2001, which means that the NDCG drops after the Enron Scandal. This agrees with behavior in Figure 4.

### 6.3 Non-stationary check

We use Figure 2 to show that many users have changing preferences over time. In this section, we compare how user preferences change as a group before and after the Enron scandal. Figures 6a and 6b show that the correlation factors between user preferences change significantly after the scandal. To quantify this change, we calculate

Table 2: Average loss comparison on two subsets

|  | MS-LR | MS-ADA | OWA1 | OWA2 | OWA3 | OWA4 | timeline | MOSR |
|---|---|---|---|---|---|---|---|---|
| Loss comparison since 1999 | 13385.4 | 12155.63 | 1855.54 | 1854.6 | 1867.75 | 1845.21 | 4169.54 | 1660.73 |
| Loss comparison since the third quarter of 2001 | 2609.03 | 2601.14 | 460.14 | 432.47 | 445.7 | 438.23 | 792.74 | 404.36 |

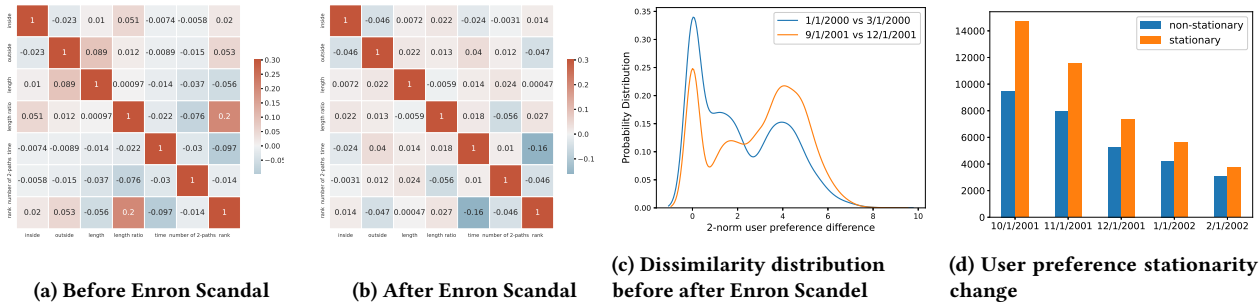

(a) Before Enron Scandal    (b) After Enron Scandal    (c) Dissimilarity distribution before after Enron Scandel    (d) User preference stationarity change

Figure 6: Comparison of users' preferences before/after Enron scandal, showing a preference shift due to external factors.

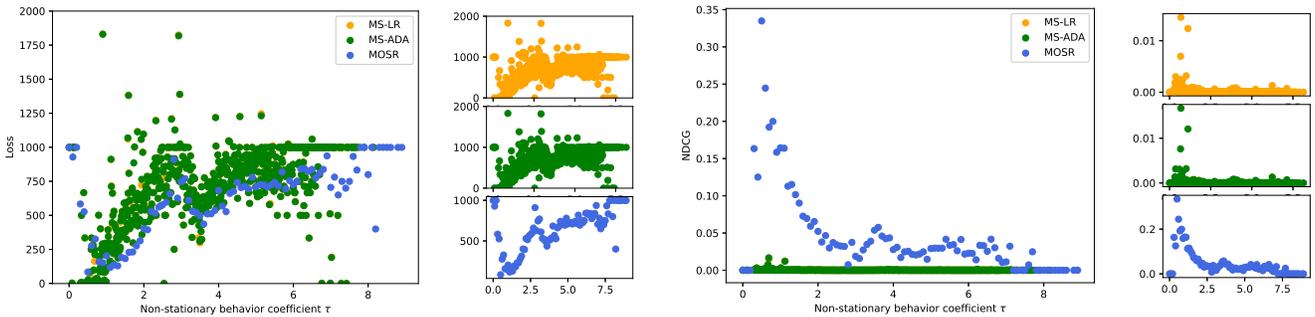

(a) Scatter plot comparing different algorithms on non-stationary behavior coefficient $\tau$ vs Loss. When the user behavior becomes more non-stationary, our loss will <u>grow slower</u> than the baselines.

(b) Scatter plot comparing different algorithms on non-stationary behavior coefficient $\tau$ vs NDCG. When the user behavior becomes more non-stationary, our NDCG will <u>drop slower</u> than the baselines.

Figure 7: Scatter plot to show how non-stationary user preferences impact the performance. Each subfigure has an enlarged version on the right side, providing a closer look at the data.

the non-stationary behavior coefficient $\tau$ as $||A - B||_2^2$, where $A$ and $B$ are the coefficient matrices of two subsets of data. Figure 6c shows how $\tau$ changes between two pairs of dates: 01/01/00 vs 03/01/00 and 09/01/01 vs 12/01/01. We see that $\tau$ increases after the scandal, which means that user preferences are more affected by external events like the scandal. To separate users with stable and unstable preferences, we set a threshold of $\tau = 1$. Users with $\tau < 1$ have stable preferences, while users with $\tau \geq 1$ have unstable ones. Figure 6d shows that there are more users with unstable preferences after the scandal.

Our algorithm outperforms the baselines because it can adapt to unstable user preferences better than they can. The loss and NDCG metrics reflect how well our algorithm matches user preferences. Higher loss means lower match quality, while higher NDCG means higher match quality. Figures 7a and 7b show that the loss increases and NDCG decreases as $\tau$ increases for both our algorithm and the baselines, but our algorithm has smaller changes than they do. This means that our algorithm is less affected by preference changes than they are.

To explain why MOSR is better at adapting to preference changes than the baselines, we conduct an additional experiment to study how non-stationary behavior affects loss. Figures 7a and 7b show how loss and NDCG vary with $\tau$. They also show that when $\tau$ gets larger, MOSR loss rises more slower than its competitors in Figure 7a and NDCG falls slower than its competitors in Figure 7b. This suggests that MOSR can adjust to preference changes faster than its competitors. [1]

### 6.4 Robustness check

To test the robustness of our method, we reduced the size of the dataset by random sampling. Figures 8 show how our method performed in these experiments. The results confirm that MOSR, which uses a symmetric aggregator with MRAC control, is better than the weighted sum aggregators (such as MS-LR and MS-ADA) at adapting to changing user preferences and maintaining a stable performance in the email re-ranking problem. Reducing the size of the dataset increases its variance and makes it more non-stationary. The experiments demonstrate that MOSR can effectively balance the trade-off between stability and adaptability over non-stationary behavior, resulting in a more robust performance than the baselines.

---
[1]Code is submitted at https://github.com/JYLEvangeline/MOSR

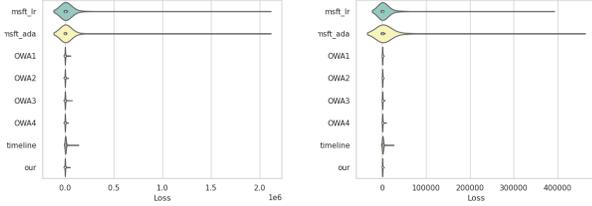

(a) Robustness check on EnronA (b) Robustness check on EnronB

Figure 8: The results of the robustness check on two down-sampling datasets indicate the stability of the algorithm, demonstrating our stationarity.

## 6.5 Parameter Tuning

As mentioned above, there are four main hyper-parameters in MOSR :

(1) **Parameters for OWA.** To generate the OWA weights, we use the RIM function with $R(x) = x^\alpha$, and we set $\alpha = 2.5$.
(2) **Parameters for MRAC learning.** For MRAC learnning, we set the time window $t_w = 3$, and learning rate $\lambda = 0.99$. We also compare $\delta_d = 0, 10, 50, 99$ and learning rate $\lambda = 0.5, 0.8, 0.9, 0.99$.
(3) **Parameters for closeness.** There are two distance measurement methods to estimate social distance $d(e_i, e_j)$ between $e_i$ and $e_j$: shortest distance and number of 2-paths.
(4) **Parameters for graph construction.** The parameters we adopt to generate graph is $k_1, k_2, k_3 = [0, 0, 2]$

We discuss two hyper-parameters: learning rate and distance measurement. The best learning rate is 0.99 and the best distance measurement is number of 2-paths. Due to the space limitation, the discussion over $\delta_d$ is in Appendix.

## 6.6 Complexity Analysis

In this section, we analyze the complexity of our algorithm versus baselines. Suppose there are $n$ email objects $G$, $m$ email addresses, and the number of features is $d$. In our model, each OWA could be regarded as an estimator. Suppose the estimators in our model and AdaBoost model are both $k$. Then the complexity is:

In Table 3, $\sigma$ refers to the complexity over feature construction. For MSFT-Logistic Regression (MS-LR) and MS-AdaBoost (MS-ADA)[30], to compute the global features HistIndiv and HistPair, the complexity is $O(m)$ and $O(m^2)$. Then, the feature construction for MS-methods are $O(\sigma) = O(m+m^2) = O(m^2)$. For simple LR and ADA models, the complexities are $O(nd^2)$ and $O(ndk)$. Hence the complexities of MS-LR and MS-ADA are $O(nd^2+\sigma)$ and $O(ndk+\sigma)$.

Table 3: Complexity Comparison

|  | Training Complexity | Updates complexity |
|---|---|---|
| MS-LR | $O(nd^2 + \sigma)$ | $O(nd^2 + \sigma)$ |
| MS-ADA | $O(ndk + \sigma)$ | $O(dk + \sigma)$ |
| OWA | $O(nd + m^2)$ or $O(nd)$ | $O(d)$ or $O(d + m^2)$ |
| timeline | $O(n)$ | $O(1)$ |
| MOSR | $O(ndk + m^2)$ or $O(ndk)$ | $O(dk + m^2)$ or $O(dk)$ |

When adding new data-point to the model, MS-LR needs to re-compute the whole model while MS-ADA only re-estimates the weights over estimators. Hence, their complexities are $O(nd^2 + \sigma)$ and $O(dk + \sigma)$, respectively.

For OWA algorithm, to compute the closeness feature, we need to calculate social distance. If we adopt shortest distance as our measurement, the complexity will be $O(\sigma) = O(m^2)$, if we adopt number of 2-paths, the complexity becomes $O(\sigma) = O(n)$. Hence, the complexity for OWA could be $O(nd + \sigma)$. If we add one data-point to OWA, we don't need to re-train the previous one, but we need to update the global features. So the complexity is $O(d + \sigma)$.

For timeline algorithm, we only compare the receiving/sending time of email entity, and decide the rankings based on time feature. So the complexity is $O(n)$.

For our model, we adopt OWA as our estimator and use MRAC to train the data. So our training complexity is $O(ndk + \sigma)$. $O(\sigma)$ is decided by the social distance option, either $O(m^2)$ or $O(n)$. When add a new data-point, the complexity is $O(dk + \sigma)$.

## 7 CONCLUSION

In this paper, we addressed the email re-ranking problem as a recommendation task based on three criteria: closeness, timeliness, and conciseness. We argued that these criteria reflect user satisfaction, and thus cannot be combined by simple weighted sums. We designed MOSR (Multi-Objective Stationary Recommender), an online algorithm that uses MRAC (Model Reference Adaptive Control) to dynamically balance the criteria and adapt to preference changes. We evaluated MOSR on the Enron Email Dataset, a large-scale real-world collection of emails, and showed that it outperforms other baselines in terms of ranking quality, especially under non-stationary preferences. We also demonstrated that MOSR is robust to high variance in email characteristics and does not require content analysis, which could raise privacy concerns. Our work contributes to the field of email re-ranking by proposing a novel method that accounts for multiple objectives impacting user satisfaction and adapts to changing user needs over time.

# A APPENDICES

## A.1 OWA change

In the background section, we discuss that the changes over parameters $\alpha$ bring the RIM quantifier to different cases. Here, we visualize the change over $\alpha$ to help readers better understand the choose of parameters in OWA. When $\alpha \to 0$, the OWA becomes MAX operator; when $\alpha \to 0$, the OWA operator becomes arithmetic mean, and when $\alpha \to \infty$, the OWA operator becomes MIN operator.

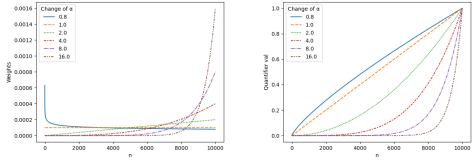

(a) Weights change with different $\alpha$   (b) Weights change with different $\alpha$

Figure 9: The OWA change with $\alpha$

## A.2 Proof for Theorem 4.2

Proof.
$$\epsilon_r = ||\mathbf{y_r} - \mathbf{y_m}||_2$$
$$= ||\sum_{i=1}^{n} \theta_i(\mathbf{y_i} - \mathbf{y_m})||_2 \quad (s.t. \sum_{i=1}^{n} \theta_i = 1) \quad (7)$$
$$\leq \sum_{i=1}^{n} ||\theta_i(\mathbf{y_i} - \mathbf{y_m})||_2 = \sum_{i=1}^{n} \theta_i \epsilon_i$$

Suppose the true ranking is $y_m$ and predicted ranking is $y_r$, in which, $y_m = \sum_{i=1}^{n} \theta_i^* y_i$ and $y_r = \sum_{i=1}^{n} \theta_i^{(j)} y_i$, $y_i$ is the ranking vector of algorithm $\Omega_i$. For simplicity, we define matrix $Y = [\mathbf{y_1}, \mathbf{y_2}, ...\mathbf{y_n}]$ and $\Theta = [\theta_1, \theta_2, ...\theta_n]$, so the loss function becomes

$$\epsilon_r^{(j+1)} = ||\mathbf{y_r^{(j+1)}} - \mathbf{y_m}||_2$$
$$= ||\sum_{i=1}^{n} \theta_i^{(j+1)} \mathbf{y_i^{(j+1)}} - \mathbf{y_m}||_2$$
$$= ||\sum_{i=1}^{n} \frac{\lambda \theta_i^{(j)}}{\lambda + 1} \mathbf{y_i^{(j)}} + \frac{\mathbf{y_k^{(j)}}}{\lambda + 1} - \mathbf{y_m}||_2 \quad (\text{Here, } \epsilon_k = \min(\mathcal{E}))$$
$$\leq ||\sum_{i=1}^{n} \frac{\lambda \theta_i^{(j)}}{\lambda + 1}(\mathbf{y_i^{(j)}} - \mathbf{y_m})||_2 + ||\frac{1}{\lambda + 1}(\mathbf{y_k^{(j)}} - \mathbf{y_m})||_2$$
$$= ||\frac{\lambda}{\lambda + 1}(\mathbf{y_r^{(j)}} - \mathbf{y_m})||_2 + ||\frac{1}{\lambda + 1}(\mathbf{y_k^{(j)}} - \mathbf{y_m})||_2$$
$$< \frac{\lambda}{\lambda + 1}\epsilon_r^{(j)} + \frac{1}{\lambda + 1}\epsilon_r^{(j)} = \epsilon_r^{(j)}$$
(8)

Hence, MOSR is stable. □

## A.3 Candidates Generation

Figure 10 visualizes the effectiveness of the candidate generation process in MOSR. We have an unsuccessful attempt to add carbon copy receivers to the candidate set. Generating a more concise candidate set to cover undiscovered candidates may be a future direction for research.

## A.4 Sensitivity Analysis

We conduct sensitivity analysis on three hyper-parameters: $\delta_d$, learning rate $\lambda$, and distance measurement methods over social distance $d(e_i, e_j)$. $\delta_d$ in Equation 5 is used to measure the cover rate of the predicted candidates set, which influences the learning taste for models: When the value of $\delta_d$ is equal to 0, the loss function $y_r$ (reference) and $y_m$ (process) will only take into account their ranking order. As the value of $\delta_d$ increases, the influence of the cover rate on the candidate solutions will become more significant.

*A.4.1 $\delta_d$.* In Table 4 and 5, we present the results of our experiments on the choice of $\delta_d$. The analysis reveal that the best selection is $\delta_d = 10$, which we use as the hyper-parameter in the main paper. We also present the cumulative loss curve by date with different $\delta_d$ over EnronA and EnronB in Figure 11b, which is the full version of Figure 4 in main paper.

Table 4: Average loss comparison over EnronA

|  | $\delta_d = 0$ | $\delta_d = 10$ | $\delta_d = 50$ | $\delta_d = 99$ |
| --- | --- | --- | --- | --- |
| MS-LR | 13364.16 | 13385.40 | 13895.24 | 15446.20 |
| MS-ADA | 12134.39 | 12155.63 | 12665.46 | 14216.43 |
| OWA1 | 1866.31 | 1855.54 | 1964.94 | 2243.42 |
| OWA2 | 1857.08 | 1854.60 | 1967.81 | 2226.79 |
| OWA3 | 1856.91 | 1867.75 | 1952.33 | 2239.91 |
| OWA4 | 1861.88 | 1845.21 | 1960.13 | 2244.77 |
| Timeliness | 4167.32 | 4169.54 | 4263.84 | 4542.66 |
| Closeness | 1842.77 | 1830.62 | 1934.58 | 2244.75 |
| Conciseness | 1853.82 | 1864.10 | 1948.82 | 2235.39 |
| MOSR | 1660.20 | 1660.73 | 1771.10 | 2051.26 |

Table 5: Average loss comparison over EnronB

|  | $\delta_d = 0$ | $\delta_d = 10$ | $\delta_d = 50$ | $\delta_d = 99$ |
| --- | --- | --- | --- | --- |
| MS-LR | 2586.97 | 2609.03 | 3138.25 | 4748.21 |
| MS-ADA | 2579.09 | 2601.14 | 3130.37 | 4740.33 |
| OWA1 | 433.59 | 460.14 | 523.48 | 844.51 |
| OWA2 | 425.46 | 432.47 | 542.19 | 837.03 |
| OWA3 | 433.71 | 445.70 | 529.59 | 836.77 |
| OWA4 | 422.75 | 438.23 | 545.83 | 848.59 |
| Timeline | 787.81 | 792.74 | 896.25 | 1211.9 |
| Closeness | 423.76 | 427.58 | 530.46 | 856.15 |
| Conciseness | 438.57 | 430.66 | 528.83 | 838.88 |
| MOSR | 384.19 | 404.36 | 487.86 | 802.06 |

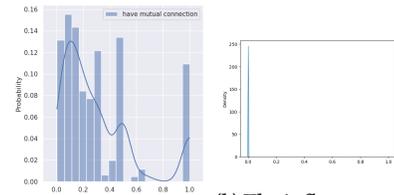

(a) The percentage of mutual connection in dates   (b) The influence over undiscovered candidates when adding undiscovered candidates carbon copy receivers to candidates set

Figure 10: The analysis for candidates set

Table 6: Tuning

|  | hyper-parameters | OWA1 | OWA2 | OWA3 | OWA4 | MOSR |
|---|---|---|---|---|---|---|
| Since 09/2001 | 0.5 | 435.09 | 436.56 | 431.88 | 435.83 | 404.14 |
|  | 0.8 | 436.2 | 450.38 | 434.5 | 441.13 | 398.13 |
|  | 0.9 | 434.47 | 424.43 | 432.3 | 420.89 | 386.13 |
|  | 0.99 | 460.14 | 432.47 | 445.7 | 438.23 | 404.36 |
|  | num of 2-paths | 460.14 | 432.47 | 445.7 | 438.23 | 404.36 |
|  | shortest path | 444.73 | 436.06 | 448.63 | 445.84 | 404.43 |
| Since 1999 | 0.5 | 1856.11 | 1871.84 | 1860.34 | 1841.78 | 1693.05 |
|  | 0.8 | 1846.55 | 1852.22 | 1875.81 | 1878.55 | 1674.44 |
|  | 0.9 | 1867.47 | 1847.95 | 1860.21 | 1895.48 | 1674.45 |
|  | 0.99 | 1855.54 | 1854.6 | 1867.75 | 1845.21 | 1660.73 |
|  | num of 2-paths | 1855.54 | 1854.6 | 1867.75 | 1845.21 | 1660.73 |
|  | shortest path | 1844.29 | 1861.4 | 1891.55 | 1863.01 | 1685.91 |

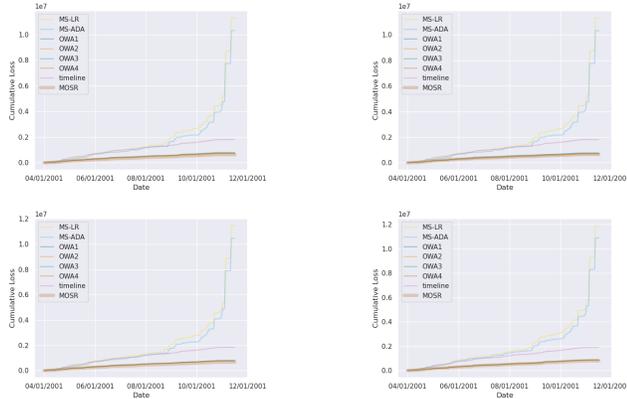

(a) The cumulative loss over EnronA with $\delta = 0, 10, 50, 99$

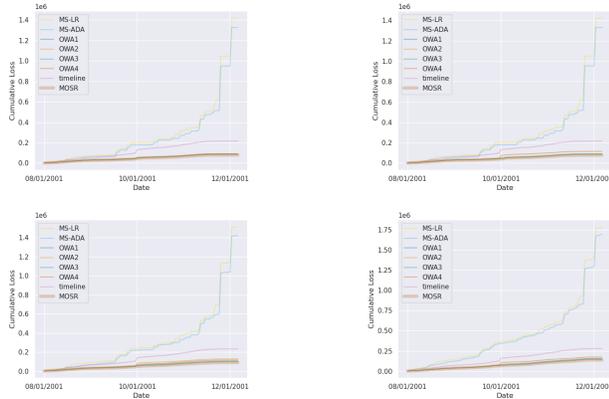

(b) The cumulative loss over EnronB with $\delta = 0, 10, 50, 99$

Figure 11: Cumulative loss curve by date with different $\delta_d$ over EnronA and Enron B

A.4.2 $\lambda$ *and distance measurement methods.* In this section, we discuss two important hyper-parameters: $\lambda$ and the distance measurement method. $\lambda$ serves as the learning rate in MRAC methods, while the distance measurement method is used to estimate the social distance, $d(e_i, e_j)$, between two users. The distance measurement method is crucial as it directly relates to the key objective of our approach, which is the calculation of closeness.

The parameter tuning results are in Table 6. Based on the tunning results, we choose $\lambda = 0.99$ and distance measurement method as number of 2-paths.

### A.5 Robustness Check

In this section, we present a comprehensive analysis of the robustness of our approach. Figure 12 clearly demonstrates that the results for OWAs are more stable than those for MSFTs. Additionally, our algorithm, MOSR , exhibits even greater stability compared to other OWA-related results. These results align with our conclusions that:

(1) The email re-ranking problem involves conflicting satisfaction criteria, which makes weighted sum aggregators an ineffective solution.
(2) The use of an adaptive control model can further enhance the stability of OWA algorithms.

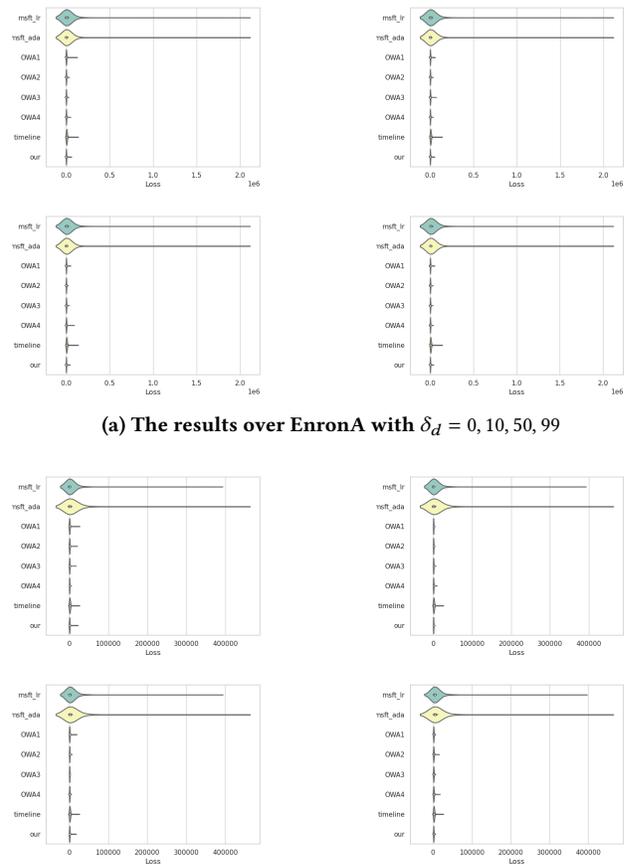

(a) The results over EnronA with $\delta_d = 0, 10, 50, 99$

(b) The results over EnronB with $\delta_d = 0, 10, 50, 99$

Figure 12: Robustness check